# ANALYSIS OF CUBIC PERMUTATION POLYNOMIALS FOR TURBO CODES

Lucian Trifina, Teaching assistant PhD. Eng.

*Technical University "Gh. Asachi" Iasi, Faculty of Electronics, Telecommunications and Information Technology Bd. Carol I no. 11, 700506, Romania.*

Phone: 040 – 232 – 213737 ext 279

Fax: 040 – 0232 – 217720

email: luciant@etti.tuiasi.ro

Daniela Tarniceriu, Professor PhD. Eng.

*Technical University "Gh. Asachi" Iasi, Faculty of Electronics, Telecommunications and Information Technology Bd. Carol I no. 11, 700506, Romania.*

Phone: 040 – 232 – 213737 ext 235

Fax: 040 – 0232 – 217720

email: tarniced@etti.tuiasi.ro

**Abstract** – Quadratic permutation polynomials (QPPs) have been widely studied and used as interleavers in turbo codes. However, less attention has been given to cubic permutation polynomials (CPPs). This paper proves a theorem which states sufficient and necessary conditions for a cubic permutation polynomial to be a null permutation polynomial. The result is used to reduce the search complexity of CPP interleavers for short lengths (multiples of 8, between 40 and 352), by improving the distance spectrum over the set of polynomials with the largest spreading factor. The comparison with QPP interleavers is made in terms of search complexity and upper bounds of the bit error rate (BER) and frame error rate (FER) for AWGN and for independent fading Rayleigh channels. Cubic permutation polynomials leading to better performance than quadratic permutation polynomials are found for some lengths.

# 1. Introduction

The permutation polynomial (PP) based interleavers are the most recent published in the literature and they are characterized by [1]: complete algebraic structure, efficient implementation (high speed and low memory requirements) and very good performances concerning bit error rates (BER) and frame error rates (FER).

The most studied PP based interleavers are those based on quadratic permutation polynomials (QPPs) [1-6]. They are used in the Long Term Evolution (LTE) standard [7].

Necessary and sufficient conditions for equivalence of QPP based interleavers have been established in [8-9]. Independently, equivalence conditions, but just for sufficiency, were given in [10]. These conditions are useful to reduce the complexity of search QPP based interleavers, when used in turbo codes.

Less attention was paid to cubic permutation polynomials (CPPs) [11-12]. In this paper we want to find equivalence conditions for CPP based interleavers. They are given by means of the null permutation polynomial (NPP) [9], i.e. a permutation polynomial equal to zero for all the elements to be permuted. In addition, we want to look for CPPs appropriate for classic turbo codes with the global coding rate of 1/3 and the generator matrix with the component code G = [1, 15/13] (octal form), as in the LTE standard.

The paper is structured as follows. The second section gives some basic definitions as well as the conditions on QPP equivalence. The third section states sufficient and necessary conditions for a CPP to be cubic null permutation polynomials (CNPP) and makes an analysis in terms of search complexity of QPP and CPP based interleavers. Section 4 proposes a method to obtain QPP and CPP based interleavers leading to the largest spread and the best distance spectrum for lengths between 40 and 352, multiples of 8. Section 5 presents simulations for two interleaver lengths and the above mentioned component code of memory 3. Section 6 concludes the paper.

# 2. Basic definitions and previous results on QPP equivalence

A permutation polynomial of degree $n$ is defined as [1]:

$$\pi(x) = \sum_{k=0}^{n} q_k x^k \mod L, \ x = 0, 1, \ldots, L-1 \qquad (1)$$

where the coefficients $q_k, k = 1, \ldots, n$ are chosen so that $\pi(x)$ in (1) permutes $0, 1, \ldots, L-1$ and $q_0$ only determines a shift of the permutation elements. The permutation function is $\pi : \mathbb{Z}_L \to \mathbb{Z}_L$, where $\mathbb{Z}_L = \{0, 1, \ldots, L-1\}$.

A QPP based interleaver of length $L$ results from (1) for $n = 2$.

In the following we only consider QPPs with the free term $q_0 = 0$, as for the QPP interleavers in the LTE standard.

A CPP based interleaver of length $L$ results from (1) for $n = 3$.

The conditions that coefficients of a CPP have to fulfill so that they generate a permutation polynomial were given in [11-12].

A NPP is characterized by:

$$\pi(x) = 0 \bmod L, \qquad x = 0, 1, \ldots, L-1 \qquad (2)$$

A NPP is useful because by adding it to a given permutation polynomial we obtain another polynomial leading to an identical permutation (excepting the NPP with all null coefficients). This avoids re-calculating the parameters imposed by the search.

Null permutation polynomials have been studied in [9] and the necessary and sufficient conditions for a QPP to be NPP was given in [8]. The sufficiency condition has been independently given and it has been demonstrated in [10] in two ways, one of them being different from the classic one in [8]. Therefore, a QPP is a null permutation polynomial (QNPP), different from the polynomial with all coefficients 0, if and only if the length $L$ is an even number and

$$q_0 = 0 \text{ and } q_1 = q_2 = L/2 \qquad (3)$$

These conditions reduce the search time, avoiding the calculation of the distance spectrum for turbo codes for identical permutation polynomials. In this paper we intend to find the necessary and sufficient conditions for a CPP to be NPP. The QPP and CPP based interleavers are optimized considering the distance spectrum and they are searched in a reduced set of polynomials (those with maximum spreading factor).

The spreading factor or parameter D of an interleaver is defined by [1]

$$D = \min_{\substack{i \neq j \\ i, j \in \mathbb{Z}_L}} \{\delta_L(p_i, p_j)\}, \qquad (4)$$

where $\delta_L(p_i, p_j)$ is the Lee metric between the points $p_i = (i, \pi(i))$ and $p_j = (j, \pi(j))$:

$$\delta_L(p_i, p_j) = |i - j|_L + |\pi(i) - \pi(j)|_L, \qquad (5)$$

where

$$|i - j|_L = \min\{(i - j)(\bmod L), (j - i)(\bmod L)\}. \qquad (6)$$

The QPP based interleavers which lead to the largest spreading factor D for some lengths are given in [1]. An algorithm for faster computation of D by means of representatives of orbits in the representation of the interleaver-code is also presented. The interleaver-code is a graphic representation of an interleaver with a point for each pair $(i, \pi(i))$, $i = 0, 1, \ldots, L-1$. An orbit is a set of points, equivalent under the action of an isometry group.

## 3. Necessary and sufficient conditions for CNPP

Firstly we state a lemma which helps to prove the next theorem.
*Lemma.* The sum of the first $n$ natural numbers is:

$$\frac{n(n+1)}{2} = 3m, \text{ if } n = 3p \text{ or } n = 3p+2, \text{ with } m, p \in \mathbb{N} \tag{7}$$

or

$$\frac{n(n+1)}{2} = 3m+1, \text{ if } n = 3p+1, \text{ with } m, p \in \mathbb{N} \tag{8}$$

*Proof.*

If $n = 3p$, we have

$$\frac{n(n+1)}{2} = \frac{3p(3p+1)}{2} = 3m, \ m \in \mathbb{N} \tag{9}$$

because $p(3p+1)$ is an even number.

If $n = 3p+2$, we have

$$\frac{n(n+1)}{2} = \frac{3(p+1)(3p+2)}{2} = 3m, \ m \in \mathbb{N} \tag{10}$$

because $(p+1)(3p+2)$ is an even number.

If $n = 3p+1$, we have

$$\frac{n(n+1)}{2} = \frac{9p(p+1)}{2} + 1 = 3m+1, \ m \in \mathbb{N} \tag{11}$$

because $p(p+1)$ is an even number. □

The following theorem specifies the conditions on the interleaver length and on the coefficient values of a CPP, so that it is a CNPP.

*Theorem.* A CPP of length $L$ defined by the permutation
$$\pi(x) = q_0 + q_1 x + q_2 x^2 + q_3 x^3 \bmod L, \ x = 0, 1, \ldots, L-1 \tag{12}$$

with $q_0 = 0$ is a CNPP (with at least one nonzero coefficient), if and only if its coefficients are as in the following ten cases (I – X), under three conditions on the interleaver length (a-c):

a) if $2 | L$ and $3 \nmid L$

I) $q_1 = \frac{L}{2}$, $q_2 = 0$, $q_3 = \frac{L}{2}$

II) $q_1 = 0$, $q_2 = \frac{L}{2}$, $q_3 = \frac{L}{2}$,

b) if $3 | L$ and $2 \nmid L$

III) $q_1 = \frac{2L}{3}$, $q_2 = 0$, $q_3 = \frac{L}{3}$

IV) $q_1 = \dfrac{L}{3}$, $q_2 = 0$, $q_3 = \dfrac{2L}{3}$,

c) if $6 | L$

cases (I-IV) or

V) $q_1 = \dfrac{5L}{6}$, $q_2 = 0$, $q_3 = \dfrac{L}{6}$

VI) $q_1 = \dfrac{L}{3}$, $q_2 = \dfrac{L}{2}$, $q_3 = \dfrac{L}{6}$

VII) $q_1 = \dfrac{L}{6}$, $q_2 = \dfrac{L}{2}$, $q_3 = \dfrac{L}{3}$

VIII) $q_1 = \dfrac{5L}{6}$, $q_2 = \dfrac{L}{2}$, $q_3 = \dfrac{2L}{3}$

IX) $q_1 = \dfrac{L}{6}$, $q_2 = 0$, $q_3 = \dfrac{5L}{6}$

X) $q_1 = \dfrac{2L}{3}$, $q_2 = \dfrac{L}{2}$, $q_3 = \dfrac{5L}{6}$,

where the notation $a | b$ means that $a$ divides $b$, the notation $a \nmid b$ means that $a$ does not divide $b$.

*Proof.*

The proof of the theorem is based on the idea in [8] used for QNPP.

For a CNPP with $q_0 = 0$, we must have

$$q_3 n^3 + q_2 n^2 + q_1 n = 0 \bmod L, \ n = 1, 2, \ldots, L-1 \qquad (13)$$

Summing the relations in (13) for the first $n$ natural numbers, we have:

$$q_3 \sum_{k=1}^{n} k^3 + q_2 \sum_{k=1}^{n} k^2 + q_1 \sum_{k=1}^{n} k = 0 \bmod L, \ n = 1, 2, \ldots, L-1 \qquad (14)$$

Relation (14) can be equivalently written as:

$$q_3 \left[\dfrac{n(n+1)}{2}\right]^2 + q_2 \dfrac{n(n+1)(2n+1)}{6} + q_1 \dfrac{n(n+1)}{2} = 0 \bmod L, \ n = 1, 2, \ldots, L-1 \quad (15)$$

or

$$\dfrac{n(n+1)}{2}\left[q_3 \dfrac{n(n+1)}{2} + q_2 \dfrac{2n+1}{3} + q_1\right] = 0 \bmod L, \ n = 1, 2, \ldots, L-1 \qquad (16)$$

In the following we prove the sufficiency for each of the cases I to X.

a) Cases I and II follow directly, because of the relations below:

$$\dfrac{L}{2}(x^3 + x) = 0 \bmod L, \ x = 0, 1, \ldots, L-1 \qquad (17)$$

$$\frac{L}{2}(x^3 + x^2) = 0 \bmod L, \quad x = 0, 1, \ldots, L-1 \tag{18}$$

They are true, due to the fact that the numbers $x(x^2+1)$ and $x^2(x+1)$ are even, $\forall x \in \mathbb{N}$.

b) For the cases III and IV, as $3|L$, we consider $L = 3r$, $r \in \mathbb{N}$. We have to check the condition in (16).
    For the case III:
- if condition (7) is fulfilled, the sum in (16) becomes

$$3m(r \cdot 3m + 2r) = 3rm(3m+2) \vdots (3r) \tag{19}$$

- if condition (8) is fulfilled, the sum in (16) becomes

$$(3m+1)[r \cdot (3m+1) + 2r] = 3r(m+1)(3m+1) \vdots (3r) \tag{20}$$

The notation $(\vdots)$ in the right hand of (19) and (20) means that the sums are divisible by $3r$, that is, by $L$ and, therefore, the condition in (16) is fulfilled.
    For the case IV:
- if condition (7) is fulfilled, the sum in (16) becomes

$$3m(2r \cdot 3m + r) = 3rm(6m+1) \vdots (3r) \tag{21}$$

- if condition (8) is fulfilled, the sum in (16) becomes

$$(3m+1)[2r \cdot (3m+1) + r] = 3r(2m+1)(3m+1) \vdots (3r) \tag{22}$$

Thus, in this case the condition in (16) is also fulfilled.

c) For cases V-X, because $6|L$, we consider $L = 6r$, $r \in \mathbb{N}$. We have to check the condition in (16). We note that if condition (8) is fulfilled, then $n = 3p+1$.
    For the case V:
- if condition (7) is fulfilled, the sum in (16) becomes

$$3m(r \cdot 3m + 5r) = 3rm(3m+5) \vdots (6r), \tag{23}$$

because $m(3m+5)$ is even.
- if condition (8) is fulfilled, the sum in (16) becomes

$$(3m+1)[r \cdot (3m+1) + 5r] = 3rm(3m+1) + 6r(3m+1) \vdots (6r), \tag{24}$$

because $m(3m+1)$ is even.
    For the case VI:
- if condition (7) is fulfilled, the sum in (16) becomes

$$3m[r \cdot 3m + r(2n+1) + 2r] = 9rm(m+1) + 6rmn \vdots (6r), \tag{25}$$

because $3|9$ and $m(m+1)$ is even.

- if condition (8) is fulfilled, the sum in (16) becomes

$$(3m+1)\left[r\cdot(3m+1)+r(2n+1)+2r\right]=6rmn+3rm(3m+5)+6r(p+1)\vdots(6r) \quad (26)$$

because $m(3m+5)$ is even.

For the case VII:
- if condition (7) is fulfilled, the sum in (16) becomes

$$3m\left[2r\cdot 3m+r(2n+1)+r\right]=6r\left(3m^2+mn+m\right)\vdots(6r), \quad (27)$$

- if condition (8) is fulfilled, the sum in (16) becomes

$$(3m+1)\left[2r\cdot(3m+1)+r(2n+1)+r\right]=6r\left[mn+3rm(m+1)+p+1\right]\vdots(6r) \quad (28)$$

For the case VIII:
- if condition (7) is fulfilled, the sum in (16) becomes

$$3m\left[4r\cdot 3m+r(2n+1)+5r\right]=6r\left(6m^2+mn+3m\right)\vdots(6r), \quad (29)$$

- if condition (8) is fulfilled, the sum in (16) becomes

$$(3m+1)\left[4r\cdot(3m+1)+r(2n+1)+5r\right]=6r\left[mn+m(6m+7)+p+2\right]\vdots(6r), \quad (30)$$

For the case IX:
- if condition (7) is fulfilled, the sum in (16) becomes

$$3m(5r\cdot 3m+r)=3rm(15m+1)\vdots(6r), \quad (31)$$

because $m(15m+1)$ is even.

- if condition (8) is fulfilled, the sum in (16) becomes

$$(3m+1)\left[5r\cdot(3m+1)+r\right]=15rm(3m+1)+6r(3m+1)\vdots(6r), \quad (32)$$

because $3|15$ and $m(3m+1)$ is even.

For the case X:
- if condition (7) is fulfilled, the sum in (16) becomes

$$3m\left[5r\cdot 3m+r(2n+1)+4r\right]=6rmn+15rm(3m+1)\vdots(6r), \quad (33)$$

because $3|15$ and $m(3m+1)$ is even.

- if condition (8) is fulfilled, the sum in (16) becomes

$$(3m+1)\left[5r\cdot(3m+1)+r(2n+1)+4r\right]=6rmn+45rm(m+1)+6r(p+2)\vdots(6r),$$

$$(34)$$

because $3|45$ and $m(m+1)$ is even.

In this way the sufficiency of the theorem is proved.

In order to prove the necessity of the theorem, we write relation (16) for $n=1$, $n=2$ and $n=3$, obtaining

$$q_3 + q_2 + q_1 = 0 \bmod L, \qquad (35)$$

$$9q_3 + 5q_2 + 3q_1 = 0 \bmod L, \qquad (36)$$

$$36q_3 + 14q_2 + 6q_1 = 0 \bmod L, \qquad (37)$$

Considering (35), relations (36) and (37) become:

$$6q_3 + 2q_2 = 0 \bmod L, \qquad (38)$$

$$30q_3 + 8q_2 = 0 \bmod L, \qquad (39)$$

Multiplying (38) by 4 and subtracting it from (39), we get:

$$6q_3 = 0 \bmod L \qquad (40)$$

Equation (40) has $\gcd(L,6)$ distinct solutions modulo $L$ [13]. They are of the form:

$$q_3 = \frac{L \cdot i}{\gcd(L,6)}, i = 0, 1, \ldots, \gcd(L,6) - 1, \qquad (41)$$

where „gcd" stands for greatest common divisor.

Considering (41), (38) becomes

$$2q_2 = 0 \bmod L, \qquad (42)$$

whose solutions are

$$q_2 = \frac{L \cdot i}{\gcd(L,2)}, i = 0, 1, \ldots, \gcd(L,2) - 1 \qquad (43)$$

The solutions for $q_1$ are obtained from (35), taking into account (41) and (43). The fact that $q_0 = 0$ results from (1) and (2) for $x = 0$. As $\gcd(L,6)$ can take the values 1, 2, 3 or 6, and $\gcd(L,2)$ can take the values 1 or 2, we see immediately that all solutions (different from zero) are those given in the theorem statement. Thus, the theorem is proven. □

The theorem allows us to evaluate the search complexity of CPP interleavers. Assuming that all the coefficients between 0 and $L-1$ are taken into account and neglecting the constant term of the PP, the complexity is of order $O(L^3)$. Given the equivalence conditions between the CPPs stated by the theorem and considering those for QPPs (relation (3)), the search complexity is reduced to $O\left(\dfrac{L^3}{4}\right)$, when $2|L$ and $3 \nmid L$, to $O\left(\dfrac{L^3}{3}\right)$, when $3|L$ and $2 \nmid L$, and to $O\left(\dfrac{L^3}{12}\right)$, when $6|L$, respectively.

The search complexity of QPPs is of order $O(L^2)$. Under the equivalence condition (3), the search complexity is reduced to $O\left(\dfrac{L^2}{2}\right)$, if $2|L$.

Comparing the search complexities of CPPs and QPPs. the CPP searching is approximately $\dfrac{L}{2}$, $\dfrac{L}{3}$, and $\dfrac{L}{6}$ times, respectively, more complex than QPP searching in each of the three cases. The complexity orders are summarized in Table 1.

**Table 1** Orders of complexity for CPP and QPP search

| Conditions on $L$ | Order of complexity for CPPs | Order of complexity for QPPs | The ratio between the order of search complexity for CPPs and QPPs |
|---|---|---|---|
| $2\|L$ and $3\nmid L$ | $O\left(\dfrac{L^3}{4}\right)$ | $O\left(\dfrac{L^2}{2}\right)$ | $\dfrac{L}{2}$ |
| $3\|L$ and $2\nmid L$ | $O\left(\dfrac{L^3}{3}\right)$ | $O(L^2)$ | $\dfrac{L}{3}$ |
| $6\|L$ | $O\left(\dfrac{L^3}{12}\right)$ | $O\left(\dfrac{L^2}{2}\right)$ | $\dfrac{L}{6}$ |

## 4. QPP and CPP interleavers of small length and improved distance spectrum

In this section we look for QPP and CPP interleavers optimized for additive white Gaussian noise (AWGN) and for Rayleigh fading channels, and compare their performances. For a quick search we restricted the interleaver lengths to multiples of 8, between 40 and 352 (the first 40 lengths of the LTE standard). The turbo code is composed from the parallel concatenation of two recursive systematic convolutional codes, with the generator matrix G = [1, 15/13] (in octal form). The trellis termination of the turbo code is as for the LTE standard (i.e. as in [14], transmitting the termination bits of the second trellis). The searches exclude polynomials reducible to linear permutation polynomials. The search method is that in [10], i.e. from the set of polynomials with the largest parameter D, those with the best distance spectra are retained. The truncated upper bounds (TUB) of BER and FER are used. For AWGN channel, the formulas are [15], [16]

$$TUB(BER) = 0.5 \cdot \sum_{i=1}^{M} \frac{w_i}{L} \cdot erfc\left(\sqrt{d_i \cdot R_c \cdot SNR}\right), \tag{44}$$

$$TUB(FER) = 0.5 \cdot \sum_{i=1}^{M} N_i \cdot erfc\left(\sqrt{d_i \cdot R_c \cdot SNR}\right) \tag{45}$$

For independent Rayleigh fading channel the formulas are [17]

$$TUB(BER) = 0.5 \cdot \sum_{i=1}^{M} \frac{w_i}{L} \cdot \left( \frac{1}{1 + R_c \cdot SNR} \right)^{d_i}, \qquad (46)$$

$$TUB(FER) = 0.5 \cdot \sum_{i=1}^{M} N_i \cdot \left( \frac{1}{1 + R_c \cdot SNR} \right)^{d_i}, \qquad (47)$$

where $M$ is the number of terms from the distance spectrum taken into account, $d_i$ is the distance $i$ in the spectrum, $w_i$ is the total information weight corresponding to distance $i$, $N_i$ is the number of code words with distance $d_i$, $R_c$ is the encoding rate and SNR is the signal to noise ratio.

We minimized TUB(BER) for the AWGN channel and TUB(FER) for the Rayleigh fading channel, as this channel type is widely encountered in wireless communications, for which FER presents more interest. The obtained polynomials are denoted by LS-QPP-TUB(BER)min, LS-CPP-TUB(BER)min, LS-QPP-TUB(FER)min and LS-CPP-TUB(FER)min, respectively (LS stands for "largest spread"). To calculate the distance spectrum we have used Garello's method [18], [19]. The value of the parameter wu_max is equal to 10, as recommended in [19]. To reduce the computing time, the number of terms in the spectrum is firstly reduced when the length is greater than or equal to 120, then when it is greater or equal to 296.

The considered SNR values were decreased in (44) – (47), when the interleaver length increased, in order not to determine too small values for TUB(BER) or TUB(FER), as in [10]. Since the turbo code uses a component code with memory 3, the coding rate is calculated with the relation:

$$R_c = \frac{L}{3 \cdot L + 12} \qquad (48)$$

Table 2 gives the QPPs and CPPs found out by optimizing the distance spectrum for AWGN channel. The search for CPPs included QPPs in order not to result in a weaker interleaver than that based on QPP in terms of TUB(BER)/TUB(FER) performances. This is why in the tables we indicate LS-CPP-TUB(BER)min or LS-QPP-TUB(BER)min. For the specified SNR values and the considered number of distances, the values $10^7 \times$ TUB(BER) and $10^5 \times$ TUB(FER) are given. The value of the parameter D for each interleaver is also given, as well as the number of QPPs and CPPs which led to the highest value of D and the minimum TUB(BER) for that length. The CPP count also includes the QPPs to which the CPPs are reducible, when the largest parameter D is the same for QPPs and CPPs. The table only presents the polynomials with the lowest $q_1$, then with the lowest $q_2$ (for QPP) and then with the lowest $q_3$ (for CPPs). In the last column the ratio between the TUB(BER) for QPP and CPP interleavers is given. We observe values greater than or equal to 2 for some lengths (40, 48, 64, 72, 120), expecting better performance for these lengths.

For CPP interleavers the maximum value of the parameter D can be higher than that for QPP ones, but the performance is not necessarily better (eg. for lengths of 200, 256 and 304). For a proper comparison more extensive searches of CPPs have been made, imposing a minimum parameter D equal to the maximum one resulted for QPP interleavers, i.e. for lengths 120, 200, 240, 256, 304 and 336 (although in the case of the lengths 120 and 336 a better performance resulted). The obtained polynomials are denoted by $D_{min\text{-}imposed\text{-}LS\text{-}QPP}$-CPP-TUB(BER)min when $q_3 \neq 0$. Otherwise, a LS-QPP-TUB(BER)min polynomial results.

G=[1, 15/13]

**Table 2** LS- QPP- TUB(BER)min and LS-CPP- TUB(BER)min interleavers for AWGN channel

| $L$ | SNR [dB] | num dist | LS-QPP- TUB(BER)min Interleavers | | | | | LS-CPP- TUB(BER) min or LS-QPP- TUB(BER)min Interleavers | | | | | BER_QPP/ BER_CPP |
|---|---|---|---|---|---|---|---|---|---|---|---|---|---|
| | | | $\pi(x)$ | D | TUB (BER) $*10^7$ | TUB (FER) $*10^5$ | No. pol. QPP | $\pi(x)$ | D | TUB (BER) $*10^7$ | TUB (FER) $*10^5$ | No. pol. | |
| 40 | 5 | 9 | $13x+10x^2$ | 4 | 0.9336 | 0.1918 | 4 | $3x+8x^2+16x^3$ | 4 | 0.3970 | 0.0432 | 4 | 2.35 |
| 48 | 5 | 9 | $7x+36x^2$ | 6 | 0.1749 | 0.0264 | 2 | $5x+6x^2+12x^3$ | 6 | 0.0808 | 0.0156 | 24 | 2.16 |
| 56 | 5 | 9 | $3x+42x^2$ | 6 | 0.7177 | 0.2062 | 4 | $5x+14x^2+42x^3$ | 6 | 0.6923 | 0.1822 | 8 | 1.04 |
| 64 | 5 | 9 | $7x+16x^2$ | 8 | 0.2298 | 0.0739 | 4 | $5x+24x^2+48x^3$ | 8 | 0.0183 | 0.0062 | 8 | 12.56 |
| 72 | 4.5 | 9 | $5x+60x^2$ | 8 | 1.3881 | 0.4780 | 4 | $7x+4x^3$ | 8 | 0.0598 | 0.0169 | 12 | 23.21 |
| 80 | 4.5 | 9 | $11x+20x^2$ | 10 | 0.0131 | 0.0034 | 4 | $11x+20x^2$ | 10 | 0.0131 | 0.0034 | 8 | 1.00 |
| 88 | 4 | 9 | $5x+22x^2$ | 8 | 0.3257 | 0.1557 | 4 | $27x+22x^2+66x^3$ | 8 | 0.2544 | 0.1122 | 8 | 1.28 |
| 96 | 4 | 9 | $13x+72x^2$ | 12 | 0.1441 | 0.0612 | 4 | $9x+12x^2+56x^3$ | 12 | 0.1014 | 0.0612 | 24 | 1.42 |
| 104 | 3.75 | 9 | $7x+78x^2$ | 8 | 0.1727 | 0.0959 | 4 | $37x+78x^3$ | 8 | 0.1562 | 0.0600 | 8 | 1.11 |
| 112 | 3.5 | 9 | $41x+28x^2$ | 14 | 0.4836 | 0.2701 | 4 | $41x+28x^3$ | 14 | 0.3701 | 0.2172 | 8 | 1.31 |
| 120 | 3.5 | 7 | $17x+90x^2$ | 10 | 0.3141 | 0.1568 | 4 | $5x+48x^3$ | 12 | 0.0832 | 0.0390 | 12 | 3.78 |
| 128 | 3.5 | 7 | $17x+32x^2$ | 16 | 0.0704 | 0.0463 | 4 | $17x+32x^2$ | 16 | 0.0704 | 0.0463 | 8 | 1.00 |
| 136 | 3.5 | 7 | $19x+102x^2$ | 10 | 0.2875 | 0.1703 | 4 | $19x+34x^2$ | 10 | 0.2246 | 0.1329 | 8 | 1.28 |
| 144 | 3.25 | 7 | $19x+36x^2$ | 16 | 0.0472 | 0.0318 | 4 | $19x+36x^2$ | 16 | 0.0472 | 0.0318 | 24 | 1.00 |
| 152 | 3.25 | 7 | $59x+38x^2$ | 12 | 0.4973 | 0.3280 | 4 | $59x+114x^3$ | 12 | 0.4103 | 0.2711 | 8 | 1.21 |
| 160 | 3.25 | 7 | $19x+120x^2$ | 16 | 0.0521 | 0.0498 | 4 | $19x+40x^2+40x^3$ | 16 | 0.0469 | 0.0385 | 8 | 1.11 |
| 168 | 3 | 7 | $61x+126x^2$ | 12 | 0.7835 | 0.6563 | 4 | $3x+42x^2+154x^3$ | 12 | 0.4852 | 0.3567 | 24 | 1.61 |
| 176 | 3 | 7 | $65x+44x^2$ | 16 | 0.0963 | 0.0931 | 4 | $21x+132x^3$ | 16 | 0.0953 | 0.0861 | 4 | 1.01 |
| 184 | 3 | 7 | $25x+46x^2$ | 14 | 0.0545 | 0.0479 | 4 | $25x+46x^2$ | 14 | 0.0545 | 0.0479 | 8 | 1.00 |
| 192 | 3 | 7 | $23x+144x^2$ | 16 | 0.0265 | 0.0247 | 4 | $7x+48x^2+64x^3$ | 16 | 0.0265 | 0.0247 | 24 | 1.00 |
| 200 | 3 | 7 | $13x+150x^2$ | 14 | 0.0709 | 0.0568 | 4 | $41x+40x^2+180x^3$ | 20 | 0.0769 | 0.0808 | 8 | 0.92 |
| 208 | 2.75 | 7 | $27x+52x^2$ | 16 | 0.0528 | 0.0838 | 4 | $85x+26x^2+52x^3$ | 16 | 0.0343 | 0.0292 | 8 | 1.54 |
| 216 | 2.75 | 7 | $23x+144x^2$ | 18 | 0.0322 | 0.0359 | 4 | $11x+36x^2+144x^3$ | 18 | 0.0322 | 0.0359 | 24 | 1.00 |
| 224 | 2.75 | 7 | $27x+168x^2$ | 16 | 0.8377 | 0.9411 | 4 | $27x+56x^2+112x^3$ | 16 | 0.8377 | 0.9411 | 8 | 1.00 |
| 232 | 2.75 | 7 | $85x+58x^2$ | 16 | 0.0082 | 0.0105 | 4 | $85x+58x^2$ | 16 | 0.0082 | 0.0105 | 8 | 1.00 |
| 240 | 2.75 | 7 | $91x+60x^2$ | 16 | 0.0326 | 0.0704 | 4 | $29x+30x^2+20x^3$ | 18 | 0.0279 | 0.0371 | 24 | 1.17 |
| 248 | 2.75 | 7 | $33x+186x^2$ | 18 | 0.0119 | 0.0142 | 4 | $33x+62x^2+124x^3$ | 18 | 0.0119 | 0.0142 | 8 | 1.00 |
| 256 | 2.5 | 7 | $31x+64x^2$ | 16 | 0.0144 | 0.0139 | 4 | $19x+96x^2+192^3$ | 18 | 0.3217 | 0.4015 | 8 | 0.04 |
| 264 | 2.5 | 7 | $17x+66x^2$ | 18 | 0.0387 | 0.0521 | 4 | $17x+66x^2$ | 18 | 0.0387 | 0.0521 | 24 | 1.00 |
| 272 | 2.5 | 7 | $101x+204x^2$ | 16 | 0.0059 | 0.0069 | 2 | $101x+204x^2$ | 16 | 0.0059 | 0.0069 | 4 | 1.00 |
| 280 | 2.5 | 7 | $17x+210x^2$ | 20 | 1.8347 | 2.5452 | 4 | $17x+210x^2$ | 20 | 1.8347 | 2.5452 | 8 | 1.00 |
| 288 | 2.5 | 7 | $55x+72x^2$ | 18 | 0.0216 | 0.0198 | 4 | $55x+72x^2$ | 18 | 0.0216 | 0.0198 | 24 | 1.00 |
| 296 | 2.5 | 5 | $109x+74x^2$ | 20 | 0.0150 | 0.0266 | 4 | $109x+74x^2$ | 20 | 0.0150 | 0.0266 | 8 | 1.00 |
| 304 | 2.5 | 5 | $113x+76x^2$ | 16 | 0.0031 | 0.0027 | 4 | $47x+38x^2+76x^3$ | 18 | 0.1371 | 0.2009 | 8 | 0.02 |
| 312 | 2.5 | 5 | $19x+78x^2$ | 22 | 0.0244 | 0.0415 | 4 | $19x+78x^2$ | 22 | 0.0244 | 0.0415 | 24 | 1.00 |
| 320 | 2.25 | 5 | $21x+80x^2$ | 20 | 0.0111 | 0.0152 | 4 | $21x+80x^2$ | 20 | 0.0111 | 0.0152 | 8 | 1.00 |
| 328 | 2.25 | 5 | $39x+246x^2$ | 22 | 0.0084 | 0.0131 | 4 | $39x+246x^2$ | 22 | 0.0084 | 0.0131 | 8 | 1.00 |
| 336 | 2.25 | 5 | $125x+252x^2$ | 16 | 0.0425 | 0.0661 | 2 | $31x+126x^2+28^3$ | 18 | 0.0404 | 0.0714 | 24 | 1.05 |
| 344 | 2.25 | 5 | $21x+258x^2$ | 24 | 0.0084 | 0.0165 | 4 | $21x+258x^2$ | 24 | 0.0084 | 0.0165 | 8 | 1.00 |
| 352 | 2.25 | 5 | $153x+264x^2$ | 22 | 0.0104 | 0.0132 | 2 | $153x+264x^2$ | 22 | 0.0104 | 0.0132 | 4 | 1.00 |

Extended search results are given in Table 3 for AWGN channel. As shown, for the lengths 120, 240 and 336, the same polynomials have resulted. For the length 200 a slightly better polynomial resulted, and for the lengths 256 and 304, we obtained even the LS-QPP-TUB(BER)min. However, the search complexity increased.

**Table 3** LS-QPP- TUB(BER)min and $D_{\text{imposed-LS-QPP}}$-CPP- TUB(BER) min interleavers for AWGN channel (more extensive search)

| | | | LS-QPP- TUB(BER)min Interleavers | | | | | $D_{\text{min-imposed-LS-QPP}}$-CPP- TUB(BER) min or LS-QPP- TUB(BER)min Interleavers | | | | | |
|---|---|---|---|---|---|---|---|---|---|---|---|---|---|
| $L$ | SNR [dB] | num dist | $\pi(x)$ | D | TUB (BER) $*10^7$ | TUB (FER) $*10^5$ | No. pol. QPP | $\pi(x)$ | D | TUB (BER) $*10^7$ | TUB (FER) $*10^5$ | No. pol. | BER_QPP/ BER_CPP |
| 120 | 3.5 | 7 | $17x+90x^2$ | 10 | 0.3141 | 0.1568 | 4 | $5x+48x^3$ | 12 | 0.0832 | 0.0390 | 12 | 3.78 |
| 200 | 3 | 7 | $13x+150x^2$ | 14 | 0.0709 | 0.0568 | 4 | $3x+80x^3$ | 14 | 0.0459 | 0.0434 | 8 | 1.54 |
| 240 | 2.75 | 7 | $91x+60x^2$ | 16 | 0.0326 | 0.0704 | 4 | $29x+30x^2+20x^3$ | 18 | 0.0279 | 0.0371 | 24 | 1.17 |
| 256 | 2.5 | 7 | $31x+64x^2$ | 16 | 0.0144 | 0.0139 | 4 | $31x+64x^2$ | 16 | 0.0144 | 0.0139 | 8 | 1.00 |
| 304 | 2.5 | 5 | $113x+76x^2$ | 16 | 0.0031 | 0.0027 | 4 | $113x+76x^2$ | 16 | 0.0031 | 0.0027 | 8 | 1.00 |
| 336 | 2.25 | 5 | $125x+252x^2$ | 16 | 0.0425 | 0.0661 | 2 | $31x+126x^2+28x^3$ | 18 | 0.0404 | 0.0714 | 24 | 1.05 |

The remaining lengths lead to a TUB(BER) ratio below 2, i.e. similar performance. For lengths 80, 128, 144, 184, 192, 216, 224, 232, and those greater than 256, excepting 336, the TUB(BER) is equal to one, that is, identical interleavers from the point of view of equivalent permutation polynomial or distance spectrum.

For lengths higer than 256, excepting 336, CPP interleavers equivalent to QPP ones resulted.

However, it should be noted that for each QPP at least 2 equivalent CPPs exist (more precisely, two equivalent CPPs, when $2|L$ and $3 \nmid L$ and ten equivalent CPPs, when $6|L$). Therefore, one cannot say that a CPP is inferior to a QPP one for the specified lengths, and for the CPP and QPP classes in which the search was made, unless CPPs are irreducible to QPPs. For longer lengths it is not important for the parameter D to be the largest, but rather the metric $\Omega' = \zeta' \cdot \ln(D)$ to be the largest ($\zeta'$ is the refined nonlinearity degree and it contributes to reducing the codeword multiplicities), as shown in [1]. As a result, CPPs superior to QPPs may result, but the search complexity increases considerably for large lengths.

Table 4 presents the QPPs and CPPs obtained by optimizing the distance spectrum for Rayleigh fading channel. In the last column of Table 4 the ratio between the TUB(FER) for QPP and CPP interleavers is given. We observe values greater than or equal to 2 for four lengths (40, 64, 72, 120), the same as for the AWGN channel, excepting length 48. Identical interleavers from the point of view of equivalent permutation polynomial or distance spectrum result for the same lengths as for AWGN channel.

Table 5 provides the results when extensive searches were made, imposing the parameter D of the CPPs to be equal to or greater than that of the LS-QPP. The same polynomials as in the initial search (in Table 4) resulted for the lengths 120 and 240. For the length 200 a slightly better polynomial resulted. LS-QPP-TUB(FER)min interleavers resulted for the lengths 256, 304 and 336.

$G=[1, 15/13]$

**Table 4** LS- QPP- TUB(FER)min and LS-CPP- TUB(FER)min interleavers for Rayleigh fading channel

| | | | LS-QPP- TUB(FER)min Interleavers | | | | | LS-CPP- TUB(FER) min or LS-QPP- TUB(BER)min Interleavers | | | | | |
|---|---|---|---|---|---|---|---|---|---|---|---|---|---|
| $L$ | SNR [dB] | num dist | $\pi(x)$ | D | TUB (BER) $*10^7$ | TUB (FER) $*10^5$ | No. pol. QPP | $\pi(x)$ | D | TUB (BER) $*10^7$ | TUB (FER) $*10^5$ | No. pol. | FER_QPP/ FER_CPP |
| 40 | 7.5 | 9 | $13x+30x^2$ | 4 | 4.0451 | 0.6539 | 4 | $3x+8x^2+16x^3$ | 4 | 1.5681 | 0.1706 | 4 | 3.83 |
| 48 | 7.5 | 9 | $7x+36x^2$ | 6 | 0.7589 | 0.1150 | 2 | $5x+6x^2+12x^3$ | 6 | 0.3504 | 0.0676 | 24 | 1.70 |
| 56 | 7.5 | 9 | $3x+42x^2$ | 6 | 3.3169 | 0.9523 | 4 | $5x+14x^2+42x^3$ | 6 | 3.2002 | 0.8424 | 8 | 1.13 |
| 64 | 7.5 | 9 | $9x+48x^2$ | 8 | 1.1002 | 0.3456 | 4 | $7x+22x^2+60x^3$ | 8 | 0.1217 | 0.0233 | 8 | 14.83 |
| 72 | 7.5 | 9 | $5x+60x^2$ | 8 | 1.6174 | 0.5677 | 4 | $7x+4x^3$ | 8 | 0.0399 | 0.0121 | 12 | 46.92 |
| 80 | 6.5 | 9 | $11x+20x^2$ | 10 | 0.1369 | 0.0344 | 4 | $11x+20x^2$ | 10 | 0.1369 | 0.0344 | 8 | 1.00 |
| 88 | 6.5 | 9 | $5x+22x^2$ | 8 | 0.5231 | 0.2584 | 4 | $27x+22x^2+66x^3$ | 8 | 0.3775 | 0.1798 | 8 | 1.44 |
| 96 | 6.5 | 9 | $13x+72x^2$ | 12 | 0.2173 | 0.0942 | 4 | $5x+8x^3$ | 12 | 0.1705 | 0.0771 | 24 | 1.22 |
| 104 | 6 | 9 | $37x+26x^2$ | 8 | 0.4179 | 0.2028 | 4 | $37x+78x^3$ | 8 | 0.3063 | 0.1203 | 8 | 1.69 |
| 112 | 6 | 9 | $41x+28x^2$ | 14 | 0.3613 | 0.2200 | 4 | $41x+28x^3$ | 14 | 0.2812 | 0.1825 | 8 | 1.21 |
| 120 | 6 | 7 | $17x+90x^2$ | 10 | 0.3045 | 0.1609 | 4 | $5x+48x^3$ | 12 | 0.0632 | 0.0302 | 12 | 5.33 |
| 128 | 5.5 | 7 | $17x+32x^2$ | 16 | 0.2189 | 0.1446 | 4 | $17x+32x^2$ | 16 | 0.2189 | 0.1446 | 8 | 1.00 |
| 136 | 5.5 | 7 | $19x+102x^2$ | 10 | 0.9306 | 0.5515 | 4 | $19x+34x^3$ | 10 | 0.7229 | 0.4296 | 8 | 1.28 |
| 144 | 5 | 7 | $19x+36x^2$ | 16 | 0.2131 | 0.1431 | 4 | $19x+36x^2$ | 16 | 0.2131 | 0.1431 | 24 | 1.00 |
| 152 | 5 | 7 | $59x+38x^2$ | 12 | 2.2269 | 1.4680 | 4 | $59x+114x^3$ | 12 | 1.8411 | 1.2146 | 8 | 1.21 |
| 160 | 5 | 7 | $19x+120x^2$ | 16 | 0.2383 | 0.2274 | 4 | $19x+40x^2+40x^3$ | 16 | 0.2148 | 0.1761 | 8 | 1.29 |
| 168 | 5 | 7 | $61x+126x^2$ | 12 | 1.4998 | 1.2596 | 4 | $3x+42x^2+154x^3$ | 12 | 0.8970 | 0.6758 | 24 | 1.86 |
| 176 | 5 | 7 | $21x+44x^2$ | 16 | 0.2018 | 0.1691 | 2 | $21x+44x^2$ | 16 | 0.2018 | 0.1691 | 4 | 1.00 |
| 184 | 5 | 7 | $25x+46x^2$ | 14 | 0.1083 | 0.0959 | 4 | $35x+46x^2+138x^3$ | 14 | 0.1760 | 0.0909 | 4 | 1.06 |
| 192 | 4.5 | 7 | $23x+144x^2$ | 16 | 0.1878 | 0.1735 | 4 | $23x+144x^2$ | 16 | 0.1878 | 0.1735 | 24 | 1.00 |
| 200 | 4.5 | 7 | $13x+150x^2$ | 14 | 0.4798 | 0.3839 | 4 | $41x+40x^2+180x^3$ | 20 | 0.5130 | 0.5381 | 8 | 0.71 |
| 208 | 4.5 | 7 | $25x+52x^2$ | 16 | 0.1889 | 0.1546 | 4 | $85x+26x^2+52x^3$ | 16 | 0.0983 | 0.0841 | 8 | 1.84 |
| 216 | 4.5. | 7 | $23x+144x^2$ | 18 | 0.0925 | 0.1038 | 4 | $23x+144x^2$ | 18 | 0.0925 | 0.1038 | 24 | 1.00 |
| 224 | 4.5 | 7 | $27x+168x^2$ | 16 | 2.5039 | 2.8129 | 4 | $27x+168x^2$ | 16 | 2.5039 | 2.8129 | 8 | 1.00 |
| 232 | 4.5 | 7 | $15x+174x^2$ | 16 | 0.0252 | 0.0295 | 4 | $15x+174x^2$ | 16 | 0.0252 | 0.0295 | 8 | 1.00 |
| 240 | 4.5 | 7 | $89x+60x^2$ | 16 | 0.0897 | 0.0807 | 4 | $11x+90x^2+60x^3$ | 18 | 0.1245 | 0.0794 | 12 | 1.02 |
| 248 | 4.5 | 7 | $33x+186x^2$ | 18 | 0.0336 | 0.0406 | 4 | $33x+186x^2$ | 18 | 0.0336 | 0.0406 | 8 | 1.00 |
| 256 | 4.5 | 7 | $31x+192x^2$ | 16 | 0.0131 | 0.0122 | 4 | $19x+32x^2+64x^3$ | 18 | 0.4588 | 0.5702 | 8 | 0.02 |
| 264 | 4 | 7 | $31x+66x^2$ | 18 | 0.1758 | 0.1813 | 4 | $31x+66x^2$ | 18 | 0.1758 | 0.1813 | 24 | 1.00 |
| 272 | 4 | 7 | $101x+204x^2$ | 16 | 0.0265 | 0.0310 | 2 | $101x+204x^2$ | 16 | 0.0265 | 0.0310 | 4 | 1.00 |
| 280 | 4 | 7 | $17x+210x^2$ | 20 | 8.2333 | 11.4211 | 4 | $17x+210x^2$ | 20 | 8.2333 | 11.4211 | 8 | 1.00 |
| 288 | 4 | 7 | $55x+72x^2$ | 18 | 0.0977 | 0.0895 | 4 | $55x+72x^2$ | 18 | 0.0977 | 0.0895 | 24 | 1.00 |
| 296 | 4 | 5 | $109x+222x^2$ | 20 | 0.0861 | 0.1186 | 4 | $109x+222x^2$ | 20 | 0.0861 | 0.1186 | 8 | 1.00 |
| 304 | 4 | 5 | $113x+76x^2$ | 16 | 0.0141 | 0.0122 | 4 | $47x+38x^2+76x^3$ | 18 | 0.6171 | 0.9044 | 8 | 0.01 |
| 312 | 4 | 5 | $19x+78x^2$ | 22 | 0.1109 | 0.1883 | 4 | $19x+78x^2$ | 22 | 0.1109 | 0.1883 | 24 | 1.00 |
| 320 | 4 | 5 | $21x+80x^2$ | 20 | 0.0209 | 0.0283 | 4 | $21x+80x^2$ | 20 | 0.0209 | 0.0283 | 8 | 1.00 |
| 328 | 4 | 5 | $39x+246x^2$ | 22 | 0.0150 | 0.0236 | 4 | $39x+246x^2$ | 22 | 0.0150 | 0.0236 | 8 | 1.00 |
| 336 | 3.5 | 5 | $125x+252x^2$ | 16 | 0.3215 | 0.5000 | 2 | $31x+126x^2+28x^3$ | 18 | 0.3042 | 0.5351 | 24 | 0.93 |
| 344 | 3.5 | 5 | $21x+258x^2$ | 24 | 0.0605 | 0.1181 | 4 | $21x+258x^2$ | 24 | 0.0605 | 0.1181 | 8 | 1.00 |
| 352 | 3.5 | 5 | $153x+264x^2$ | 22 | 0.0291 | 0.0381 | 2 | $153x+264x^2$ | 22 | 0.0291 | 0.0381 | 4 | 1.00 |

**Table 5** LS-QPP- TUB(FER)min and $D_{imposed\text{-}LS\text{-}QPP}$-CPP- TUB(FER) min Interleavers for fading Rayleigh channel (more extensive search)

| | | | LS-QPP- TUB(FER)min Interleavers | | | | | $D_{min\text{-}imposed\text{-}LS\text{-}QPP}$-CPP- TUB(FER) min or LS-QPP- TUB(FER)min Interleavers | | | | | |
|---|---|---|---|---|---|---|---|---|---|---|---|---|---|
| $L$ | SNR [dB] | num dist | $\pi(x)$ | D | TUB (BER) $*10^7$ | TUB (FER) $*10^5$ | No. pol. QPP | $\pi(x)$ | D | TUB (BER) $*10^7$ | TUB (FER) $*10^5$ | No. pol. | BER_QPP/ BER_CPP |
| 120 | 6 | 7 | $17x+90x^2$ | 10 | 0.3045 | 0.1609 | 4 | $5x+48x^3$ | 12 | 0.0632 | 0.0302 | 12 | 5.33 |
| 200 | 4.5 | 7 | $13x+150x^2$ | 14 | 0.4798 | 0.3839 | 4 | $3x+80x^3$ | 14 | 0.3141 | 0.2943 | 8 | 1.31 |
| 240 | 4.5 | 7 | $89x+60x^2$ | 16 | 0.0897 | 0.0807 | 4 | $11x+90x^2+60x^3$ | 18 | 0.1245 | 0.0794 | 12 | 1.02 |
| 256 | 4.5 | 7 | $31x+192x^2$ | 16 | 0.0131 | 0.0122 | 4 | $31x+192x^2$ | 16 | 0.0131 | 0.0122 | 8 | 1.00 |
| 304 | 4 | 5 | $113x+76x^2$ | 16 | 0.0141 | 0.0122 | 4 | $113x+76x^2$ | 16 | 0.0141 | 0.0122 | 8 | 1.00 |
| 336 | 3.5 | 5 | $125x+252x^2$ | 16 | 0.3215 | 0.5000 | 2 | $125x+252x^2$ | 16 | 0.3215 | 0.5000 | 12 | 1.00 |

Regarding the number of resulted polynomials, we can note the following:
- for QPP interleavers the number of polynomials is 4 or 2;
- for CPP interleavers the number of polynomials is 4 or 8, when $2|L$ and $3\nmid L$ and 12 or 24, when $6|L$.

The situation when two polynomials result for QPPs and four or twelve polynomials result for CPPs is related to the number of equivalent permutation polynomials, according to conditions given in (3) and the theorem in Section 3. Doubling these numbers in each case can be explained as follows: the inverse of both QPPs and CPPs may differ from the original ones and also have the same number of equivalent polynomials. The distance spectrum results identical for the inverse polynomial because of the construction symmetry of turbo codes (parallel concatenation of two recursive systematic convolutional codes).

Imposing the parameter D influences the number of determined polynomials. Figure 1.a shows the number of different polynomials, including LS-CPP (and LS-QPP, when the parameter D is the same for LS-QPP and LS-CPP interleavers) and LS-QPP interleavers, respectively, for which the distance spectrum is calculated with the parameter D shown in Tables 2 and 4 (i.e. the largest possible). Figure 1.b shows the ratio between the number of mentioned polynomials (LS-CPPs or LS-CPPs and LS-QPPs) and the number of LS-QPPs, depending on their length. We note that, excepting the lengths 120, 256 and 304, when the parameter D differs for interleavers based on LS-CPPs and LS-QPPs, the number of polynomials for which the distance spectrum was calculated for LS-CPPs or LS-CPPs and LS-QPPs is greater than that of the LS-QPPs ones (as it was expected, when the parameter D is the same for LS-CPP and LS-QPP based interleavers).

The reasons for which for the lengths 120, 256 and 304 the number of CPPs gets lower than the number of QPPs are:
1) the largest parameter D of CPPs is higher than that of QPPs (and thus the class of CPPs with the largest parameter D does not include the class of QPPs with the largest parameter D);
2) the total number of CPPs for which the distance spectrum is calculated is reduced 2 or 6 times, respectively, compared to the total number of QPPs (as shown in Table 1).

As shown in Figure 1.b the complexity of calculating the distance spectrum for LS-CPP (or LS-CPP and LS-QPP) interleavers is at most 2 to 4 times higher than for LS-QPP interleavers (with 8 exceptions from the 40 lengths) for the set with maximum parameter D, which is a substantial reduction compared to the exhaustive search (see analysis in Section 3). However, for larger lengths, the search complexity increases and also the required time, because too many polynomials with maximum parameter D result and the distance spectrum computation is more time consuming for larger lengths.

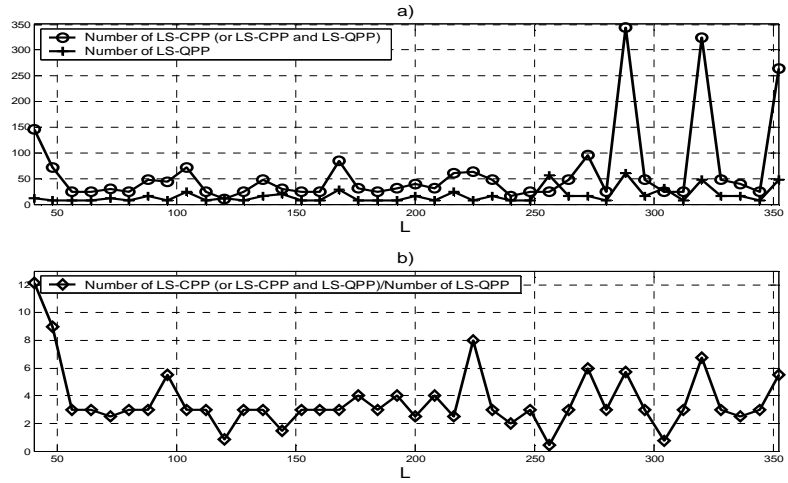

**Fig. 1** a) Number of LS-CPPs (or LS-CPPs and LS-QPPs) and LS-QPPs with the largest parameter D (given in Tables 1 and 3) for which the distance spectrum has been calculated considering the proposed optimization, b) The ratio between the numbers of LS-CPPs (or LS-CPPs and LS-QPPs) and LS-QPPs, depending on the interleaver length

Figure 2a shows the number of polynomials (Dmin-imposed-LS-QPP-CPP and LS-QPP) for the 6 lengths (120, 200, 240, 256, 304 and 336) for which the extended searches in Tables 3 and 5 were made.

Figure 2b shows the ratio between the number of the above mentioned polynomials and that of LS-QPP polynomials. This ratio reaches up to approximately 13 for the lengths 200 and 256, showing the increasing complexity of the search, compared to the previous case.

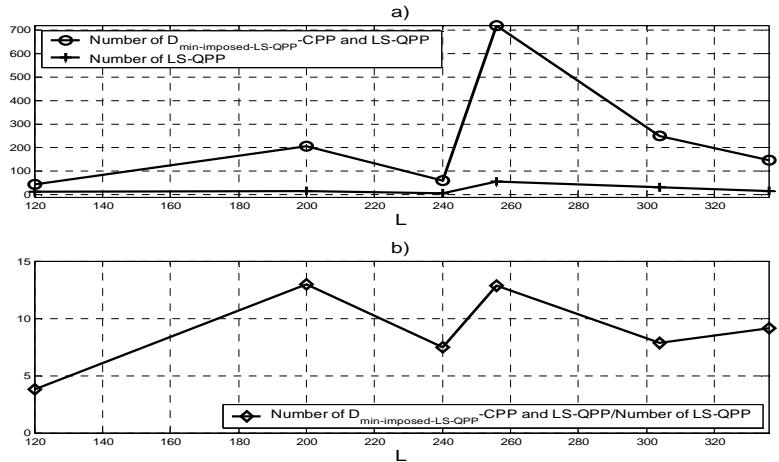

**Fig. 2** a) Number of $D_{min\text{-imposed-LS-QPP}}$-CPPs and LS-QPPs and the one of LS-QPPs (given in Tables 3 and 5) for which the distance spectrum has been calculated considering the proposed optimization, b) The ratio between the number of $D_{min\text{-}imposed\text{-}LS\text{-}QPP}$-CPPs and LS-QPPs and the number of LS-QPPs, depending on the interleaver length.

## 5. Simulation results

Simulations were performed for interleaver lengths equal to 64 and 120. The component code we considered is the one mentioned in Section 4, i.e. that corresponding to the generator matrix G = [1, 15/13]. The decoding algorithm is the MAP (Maximum-Aposteriori) criterion and the iteration stop criterion based on the LLR (Logarithm Likelihood Ratio) module. The maximum number of iterations is 12, and the LLR threshold is 10. We have simulated the same number of blocks of bits for each SNR value. Obviously, the required number of blocks increases with the SNR value. The simulated channels were AWGN and Rayleigh fading, respectively, and the used modulation BPSK (Binary Phase Shift Keying).

For both lengths we have simulated LS-CPP-TUB(BER)min and LS-QPP-TUB(BER)min interleavers for AWGN channel, and LS-CPP-TUB(FER)min and LS-QPP-TUB(FER)min interleavers for Rayleigh fading channel, respectively.

Since for the length 64 for Rayleigh fading channel the LS-CPP-TUB(FER)min and LS-QPP-TUB(FER)min interleavers lead to similar BER/FER performances, LS-CPP-TUB(BER)min and LS-QPP-TUB(BER)min were also determined and the corresponding interleavers were simulated.

Figure 3 shows BER and FER curves for length 64. We note, for AWGN channel, an additional coding gain for the LS-CPP-TUB(BER)min interleaver compared to LS-QPP-TUB(BER)min interleaver of approximately 0.25 dB in the BER domain (at BER = $10^{-6}$), and of approximately 0.5 dB in FER domain (at FER = $10^{-5}$).

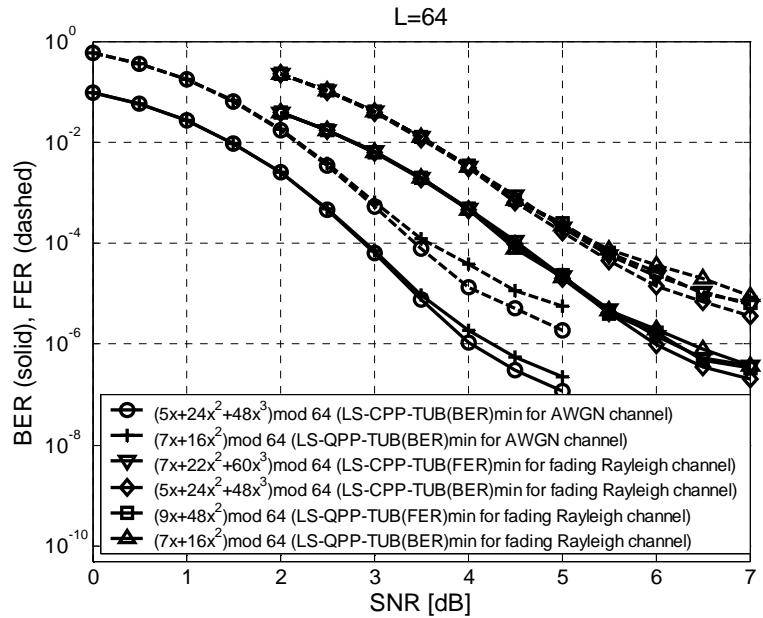

**Fig. 3** BER and FER curves for LS-CPP-TUB(BER)min and LS-QPP-TUB(BER)min interleavers for AWGN channel, and for LS-CPP-TUB(FER)min, LS-CPP-TUB(BER)min, LS-QPP-TUB(FER)min and LS-QPP-TUB(BER)min interleavers for Rayleigh fading channel, for length 64

As we have already mentioned above, for Rayleigh fading channel the LS-CPP-TUB(FER)min and the LS-QPP-TUB(FER)min interleavers lead to similar BER/FER performances. To realize why this happens, the first 20 terms of their distance spectra have been determined. The first two distances with the corresponding multiplicities for the LS-CPP-TUB(BER)min interleaver are 15/1/1 and 16/2/4. The remaining distances up to 34 are the same for the two interleavers, with higher multiplicities for the LS-CPP-TUB(FER)min interleaver (except the distance 21, when the multiplicities are 56/172 for the LS-CPP-TUB(BER)min interleaver compared to 37/137, for the LS-CPP-TUB(FER) min interleaver).

The minimum distance resulted with the LS-CPP-TUB(FER)min interleaver is greater than 16. From the distance spectra calculated for the LS-CPP-TUB(FER)min and LS-CPP-TUB(BER)min interleavers it was found that for weights greater than 16 up to 34, there are more codewords of these weights for the LS-CPP-TUB(FER)min interleaver than for the LS-CPP-TUB(BER)min interleaver (i.e., higher multiplicities). For weights less than or equal to 16, there are only three codewords of the LS-CPP-TUB(BER)min interleaver (a code word of weight 15 and two codewords of weight 16, as shown by the previously specified multiplicities: 1/1 and 2/4). The three code words have minor influence on the TUB(BER) and TUB(FER) for LS-CPP-TUB(FER)min interleaver. Considering the simulation results, we can say that there are codewords of weights greater than 34 for which the multiplicities for this interleaver are higher than those corresponding to LS-CPP-TUB(BER)min one. The higher multiplicities for weights greater than 34 lead to larger TUB(BER) and TUB(FER) for a large number of terms in distance spectrum for the LS-CPP-TUB(FER)min interleaver than for the LS-CPP-TUB(BER)min one, showing the BER/FER performances difference.

For Rayleigh fading channel the LS-CPP-TUB(BER)min interleaver leads to an additional coding gain of approximately 0.155 dB in the BER domain (at BER=$10^{-6}$) and of approximately 0.3 dB in the FER domain (at FER = $10^{-5}$), compared to the other two interleavers. The LS-QPP-TUB(BER)min interleaver leads to slightly weaker performance than LS-CPP-TUB(FER)min and LS-QPP-TUB(FER)min interleavers. These gains are lower compared to those for AWGN channel. The Rayleigh fading channel requires an additional SNR about 2 dB for the same error rate compared to AWGN channel.

BER and FER curves are given in Figure 4 for the length 120. For AWGN channel, we note an additional coding gain for the LS-CPP-TUB(BER)min interleaver compared to the LS-QPP-TUB(BER)min interleaver of approximately 0.25 dB, both in BER domain (at BER = $10^{-6}$) and in the FER domain (at FER = $10^{-5}$).

For Rayleigh fading channel we note an additional coding gain for the LS-CPP-TUB(FER)min interleaver compared to LS-QPP-TUB(FER)min interleaver of approximately 0.15 dB in the BER domain (at BER = $10^{-6}$) and of approximately 0.5 dB in the FER domain (at FER = $10^{-5}$). We also note a lower gain for the BER and higher gain for the FER compared to those for AWGN channel, due to the fact that the interleaver was searched by minimizing TUB(FER). Rayleigh fading channel again requires an additional SNR about 2 dB for the same error rate.

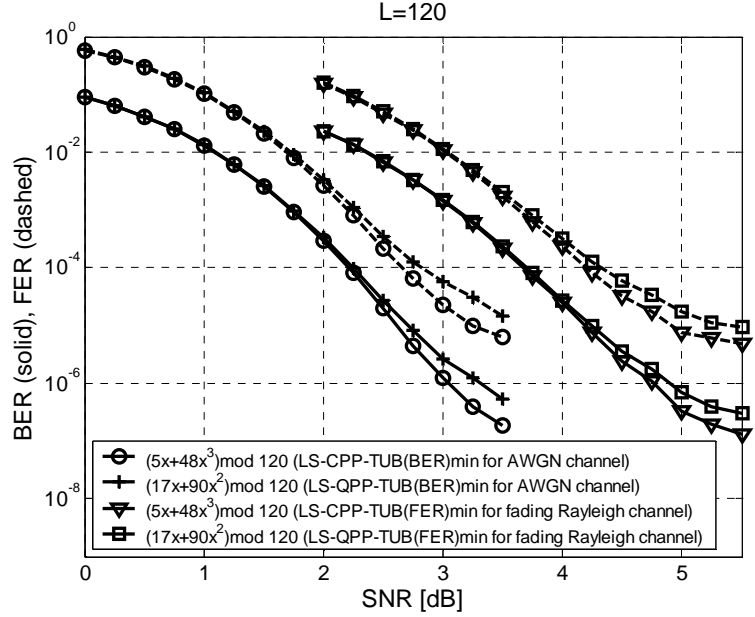

**Fig. 4** BER and FER curves for LS-CPP-TUB(BER)min and LS-QPP-TUB(BER)min interleavers for AWGN channel, and for LS-CPP-TUB(FER)min and LS-QPP-TUB(FER)min interleavers for Rayleigh fading channel, for length 120

## 6. Conclusions

The paper states and proves the necessary and sufficient conditions to be met by the coefficients of a cubic permutation polynomial to be a null permutation polynomial. The result can be used to reduce the search complexity of CPP interleavers, avoiding the distance spectrum calculation for an interleaver identical to a previous one. On the basis of this result a comparison is made in terms of searching complexity for CPP and QPP interleavers. It is shown that for CPP interleavers the complexity increases for lengths equal to $L$, of approximately $\frac{L}{2}$ times compared to that for QPP interleavers, when $2|L$ and $3 \nmid L$, of $\frac{L}{3}$ times, when $3|L$ and $2 \nmid L$, and of $\frac{L}{6}$ times, when $6|L$.

QPPs and CPPs of short lengths (multiples of 8, between 40 and 352) are searched over the restricted set of polynomials with the largest spreading factor or parameter D, by optimizing the distance spectrum for AWGN and independent Rayleigh fading channel, respectively. This reduces the search complexity, being mostly 2 to 4 times higher for CPPs than for QPPs. CPPs lead to better performance for several lengths. The simulations in Section 5 confirm this.